# Understanding Spin Configuration in the Geometrically Frustrated Magnet TbB$_4$: a Resonant Soft X-ray Scattering Study

H. Huang,[1,¶] H. Jang,[1,¶] B. Y. Kang,[2] B. K. Cho,[2] C-C Kao,[3] Y.-J. Liu,[1,†] and J.-S. Lee[1,†]

[1]*Stanford Synchrotron Radiation Lightsource, SLAC National Accelerator Laboratory, Menlo Park, California 94025, USA*

[2]*School of Materials Science and Engineering, Gwangju Institute of Science and Technology, Gwangju 61005, Korea*

[3]*SLAC National Accelerator Laboratory, Menlo Park, California 94025, USA*

[¶]These authors contributed equally to this work.

†Correspondence to: liuyijin@slac.stanford.edu and jslee@slac.stanford.edu

## Abstract:

The frustrated magnet has been regarded as a system that could be a promising host material for the quantum spin liquid (QSL). However, it is difficult to determine the spin configuration and the corresponding mechanism in this system, because of its geometrical frustration (*i.e.*, crystal structure and symmetry). Herein, we systematically investigate one of the geometrically frustrated magnets, the $TbB_4$ compound. Using resonant soft x-ray scattering (RSXS), we explored its spin configuration, as well as Tb's quadrupole. Comprehensive evaluations of the temperature and photon energy / polarization dependences of the RSXS signals reveal the mechanism of spin reorientation upon cooling down, which is the sophisticated interplay between the Tb spin and the crystal symmetry rather than its orbit (quadrupole). Our results and their implications would further shed a light on the search for possible realization of QSL.

# 1. Introduction

The quantum spin liquid (QSL) is a strongly correlated state. Unlike a conventional magnet, the QSL does not break any symmetry and remains fully dynamic at low temperature down to zero Kelvin [1-4]. Devoid of static correlations, it has been argued to support exotic fractionalized excitations called '*spinons*' [5-6]. Meanwhile, the frustrated magnet system serves as an ideal platform for the search of possible realizations of QSL. In particular, the spin-1/2 Heisenberg kagomé frustrating system, whose 2-dimensional lattice is formed by corner-sharing triangles, is one of the most promising systems for observing the QSL ground state [7-11]. Recent numerical studies of herbertsmithite $ZnCu_3(OH)_6Cl_{12}$ under magnetic field predicts a series of magnetization plateaus [12-13]. It was proposed that the plateau near zero magnetization is related to the finite gap of the putative QSL ground state, while the plateaus at 1/3, 5/9, and 7/9 correspond to different symmetry breaking magnetic orderings. Although it is still at an early stage for understanding such plateau states in frustrated magnets due to the lack of experimental evidence, it is an active research subject because it could be a promising host material for the QSL.

In the meantime, the similar plateaus phenomena have been also observed in several materials, whose lattices are topologically equivalent to the geometrically frustrated Shastry-Sutherland lattice (SSL) system [14]. Similar to the herbertsmithite compound, one of SSL systems, the $SrCu_2(BO_3)_2$, undergoes a series of fractional magnetization plateaus of the quantum spin-1/2 $Cu^{2+}$ ions [15,16]. Moreover, another SSL system, the $TbB_4$ compound, also exhibits magnetization plateaus due to strong magnetic frustration [17-20]. In $TbB_4$, a number of magnetization plateaus are observed when the magnetic field is applied along the crystalline *c*-axis, which is perpendicular to the magnetic easy-axis. It has suggested that the Tb's spin rotation in the *ab* crystal-plane is the fundamental mechanism responsible for these plateaus. The understanding of the driving force for such spin rotation could lead to valuable clues for the search of QSL, however, much remains unclear.

Here, we study $TbB_4$ compound using resonant soft x-ray scattering (RSXS) at the Tb $M_5$-edge. Although we are aware of the fact that the neutron scattering has much larger cross-section of spin [21], we employed the RSXS approach because it can directly explore the 4*f*-valence state via Tb $3d \rightarrow 4f$ dipole transition, providing valuable information regarding the spin configuration as well as the orbit (in the rear-earth case *quadrupole*) [22]. In this work, we clearly observed the antiferromagnetic order in the *ab*-plane, which turns on at $T \sim 44K$ ($T_{N1}$). Upon further cooling to below $T \sim 22$ K ($T_{N2}$), we observed the Tb spin rotation in the *ab* crystal-plane (at $T = 19$ K) with respect to the spin configuration between $T_{N1}$ and $T_{N2}$. Interestingly, our polarization dependent RSXS profile shows no change in the Tb orbital shape upon the spin rotation. This finding indicates that the Tb's spin rotation, which is regarded as the origin of the magnetization plateaus in $TbB_4$ compound [18,19], is not caused by the quadrupolar order. Instead, we found a signature of lattice distortion below $T_{N2}$ using the polarization dependence of x-ray absorption spectroscopy (XAS). Based on these findings, we conclude that the driving force of the plateaus formation in $TbB_4$ compound is associated with the structural transition (tetragonal $\rightarrow$ orthorhombic), in agreement with the speculation in Ref. [23-26].

## 2. Experiments

The TbB$_4$ single crystal subjected to our investigation was synthesized by the solution growth method using Al flux. A stoichiometric mixture of Tb (99.9%) and B (99.9%) was placed in an alumina crucible with Al (99.999%) flux with a mass ratio of TbB$_4$ : Al = 1 : 50. The mixture is melted at $T$ = 1500 °C under an argon atmosphere and slowly cooled down from 1500 °C to 650 °C with a cooling rate of 5 °C/h. TbB$_4$ has a primitive tetragonal structure (space group $P4/mbm$) with four formula units per lattice point as the basis [see Fig. 1(a)]. The grown crystals have the lattice parameters of $a$ ($b$) = 7.117 Å and $c$ = 4.028 Å [27]. The in-plane crystal structure is topologically equivalent to the SSL [17]. Figure 1(b) shows the initial spin configuration below $T_{N1}$ = 44 K, which is proposed by the previous neutron scattering experiment [23]. Also, in our previous work [28], the magnetic susceptibility measurement of the TbB$_4$ crystal shows that there is clear a second phase transition around 22 K [see Fig. 1(c)].

For the RSXS experiment, the crystal was cut and polished perpendicular to the crystalline $a$-axis to optimize the signal of the $\boldsymbol{q}$ = (1 0 0) reflection. The RSXS measurements were carried out at beamline 13-3 of Stanford Synchrotron Radiation Lightsource (SSRL) with the scattering configuration shown in Fig. 2(a). To determine the critical x-ray photon energy for the RSXS measurements, we first performed XAS measurement on the TbB$_4$ crystal. As shown in Fig. 2(b), it shows the huge enhancement at photon energy $E_{ph}$ ~ 1245 eV, which corresponds to the Tb $M_5$-edge. We then measured the scattering signals at $E_{ph}$ with either π [Fig. 2(c)] or σ [Fig. 2(d)] incident polarizations. In our scattering geometry [Fig. 2(a)], the magnetic signal is only detectable with the π polarized incident beam because the σ channel is insensitive to the

*b*-axis component of Tb magnetic moment in (1 0 0) reflection, which is indicated by the blue arrows in Fig. 2(a). According to the resonant scattering form factor [22,29,30], zero spin-cross sections in both (σ' × π) and (π' ×σ) channel are expected.

## 3. Results and discussion

Figure 2(c) shows a huge enhancement of RSXS intensity at $q$ = (1 0 0) at $T$ = 26 K (*i.e.*, below $T_{N1}$) with the π polarized illumination. This signal confirms that the in-plane noncollinear spin structure is in the antiferromagnetic (AFM) states with the spin moments, in which the spins are lying along diagonal [110] direction [see Fig. 1(b)]. Interestingly, we still detect a weak signal at the same $q$-vector above $T_{N1}$ where no-magnetic order is anticipated. As a result, it can be distinguished from the magnetic signal and is attributed to a forbidden reflection, namely an anisotropic tensor susceptibility (ATS) reflection [31-32]. Due to the crystal symmetry, most of the rare-earth tetraborides show the ATS reflection at the resonant position, which persists up to very high temperature (even over the room temperature) [31-33]. Since this signal is intrinsically similar to Jahn-Teller assisted orbital behavior in manganites [34,35], we could get information about the 4*f* orbital [22]. In this sense, we could see the similar signal in the σ incident polarization [see Fig. 2(d)]. Furthermore, we also observed the enhancement of the ATS signal below $T_{N1}$. Since the σ channel in this scattering geometry cannot detect any magnetic signal, our results suggest that the enhancement of the ATS signal is due to a quadrupolar order (QO) in Tb.

   To scrutinize more details of this polarization effect, we investigated temperature dependence of the RSXS signals at $q$ = (1 0 0) in both channels. Figure 3(a) shows the temperature dependence of AFM order in the π-polarization. Although there is finite ATS intensity above $T_{N1}$, most of the signal is properly responding to the AFM transitions. Furthermore, we clearly detect that the AFM intensity increases at $T_{N2}$, indicating that the rotation of the spins is in a direction that enhances the scattering cross-section. Considering our scattering configuration shown in Fig. 2(a), the spins are rotating toward *b*-axis direction. These transitions were also detectable by means of correlation length behavior ($\xi_{AFM} \sim 453$ Å) of the AFM order. The RSXS signal at $\boldsymbol{q} = (1\ 0\ 0)$ in the σ channel [see Fig. 3(b)], on the other hand, shows the monotonic incensement over a large temperature window up to 70 K. As we expected (*i.e.*, a development of the QO), its signal is also enhanced below $T_{N1}$. However, there is no transition around $T_{N2}$, because this signal is irrelevant to the magnetism. This temperature dependent behavior is also supported by the different correlation ($\xi_{QO} \sim 305$ Å). Accordingly, these different temperature tendencies indicate that the signals of the π and σ channels are originated from different mechanisms.

Using the resonant profiles of Tb, we further investigated the behavior of the RSXS signals in both channels. Figures 4(a) and 4(b) show the Tb resonant profile map around $\boldsymbol{q} = (1\ 0\ 0)$ of the π and σ channels, respectively. Clearly, before the AFM onset (Figure 4, top panels), the resonant profiles are same in both channels. At $T < T_{N1}$ (Figure 4, middle panels), the resonant profile in the π channel is completely changed, because the AFM order is turned on. Upon further cooling down, the π resonant profile is slightly modified around the maximum $E_{ph} \sim 1246.6$ eV, because of the Tb's spin rotation below $T_{N2}$. The σ channel's profile, on the other hand, does not show any significant difference throughout the entire temperature window [Fig. 4(b)], indicating that the signal is originated from the Tb orbital, which remain unchanged. Therefore, we conclude that the nature of the π and σ signals are presented by the spin and quadrupole, respectively. The QO remains unchanged when the second transition (*i.e.*, the spin rotation) occurs at $T_{N2}$. Our findings are schematically summarized in Fig. 5. Note that the alignment of the quadrupole moment was employed from the GdB$_4$ case [22,33] and Ref.[36].

Considering the previous works [23-26], it is possible that the TbB$_4$ undergoes a structural transition around $T_{N2}$ (*i.e.*, tetragonal → orthorhombic). Unfortunately, due to the long wavelength of the soft x-ray, RSXS is not the ideal tool for the study of the crystal structure. To gain insights in to the structural transition, we analyzed the polarization dependence of XAS spectra below and above $T_{N2}$, which leads to a linear dichroism (LD) signal that is proportional to crystal symmetry [37,38]. As shown in Fig. 6, under the grazing incident geometry (incident beam at 25$^{\mathrm{o}}$ from the sample surface), we measured *E//a* and *E//b* XAS spectra at *T* = 26 and 19.5 K. No meaningful LD (= *E//a* – *E//b*) signature at *T* = 26 K [see Fig. 6(a)] is observed, suggesting no structural difference between *a* and *b* axis. We, however, observed an LD signature at 19.5 K (*i.e.*, below $T_{N2}$) [see Fig. 6(b)]. The temperature dependence of the LD signal suggests that the second transition (*i.e.*, Tb spins' rotation) is associated with the structural transition, consistent with the previous predictions [23-26].

## 4. Conclusions

In summary, we study the spin configuration and the corresponding mechanism in the geometrically frustrated magnet TbB$_4$ using resonant soft x-ray scattering. We systematically investigated the polarization, temperature, and photon energy dependences of the RSXS in TbB$_4$. Experimentally, we found that the antiferromagnetic Tb spin's rotation happens below $T_{N2}$, which is accompanied by the structural distortion. This result provides the critical information for understanding the magnetization plateaus in TbB$_4$ compound [17-20]. Through our systematic investigation, we infer a minimized role of orbital in this case. The presented results and their implications are valuable for future efforts that search for the QSL.

## Acknowledgments

Resonant soft x-ray scattering and soft x-ray spectroscopy measurements were carried out at the Stanford Synchrotron Radiation Lightsource (SSRL), SLAC National Accelerator Laboratory, is supported by the U.S. Department of Energy, Office of Science, Office of Basic Energy Sciences under Contract No. DE-AC02-76SF00515. B. Y. K. and B. K. C were supported by National Research Foundation of Korea (NRF) funded by the Ministry of Science, ICT & Future Planning (No. NRF-2017R1A2B2008538).

# Figure

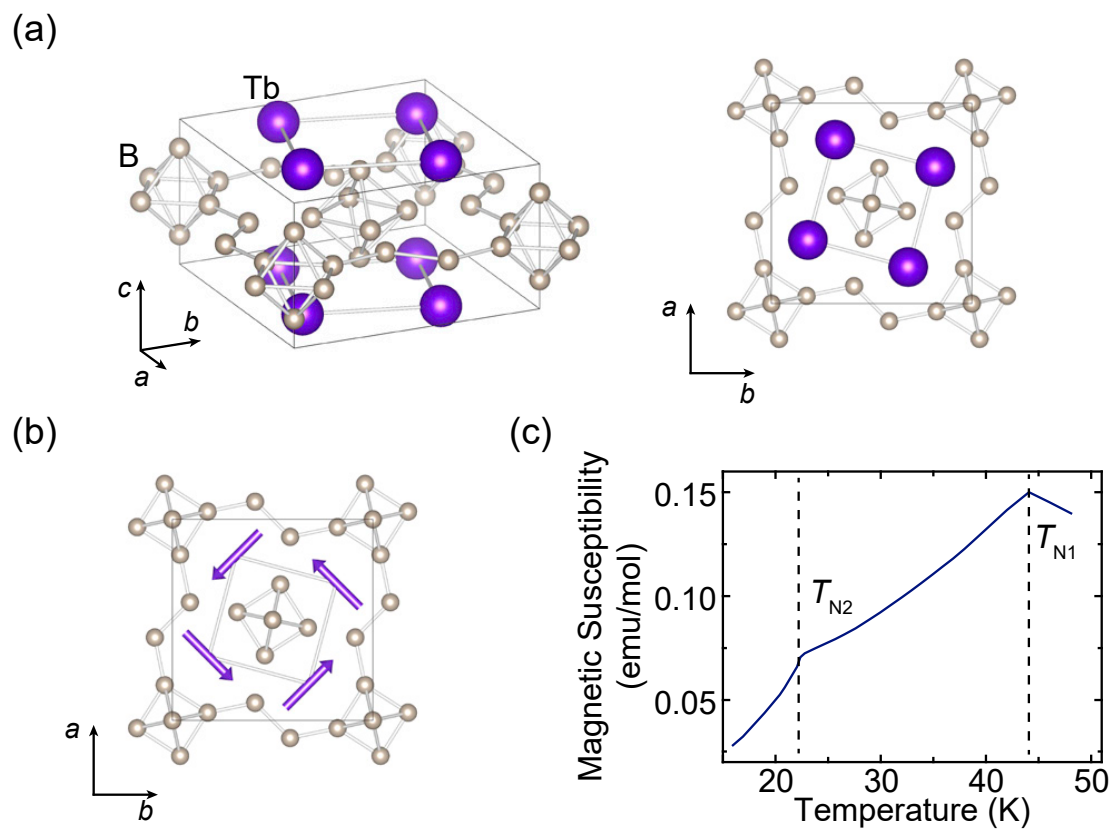

**Figure 1:** (a) Crystal structure of the geometrically frustrated magnet TbB$_4$ compound (space group *P*4/*mbm*). The left panel shows the 3-dimensional perspective shape and the right panel shows the *ab*-plane perspective shape. (b) The non-collinear spin configuration of TbB$_4$ below $T_{N1}$ in the *ab*-plane. (c) The magnetic susceptibility result of TbB$_4$ which is taken from Ref.[28]. It shows the clear transitions at both $T_{N1}$ and $T_{N2}$.

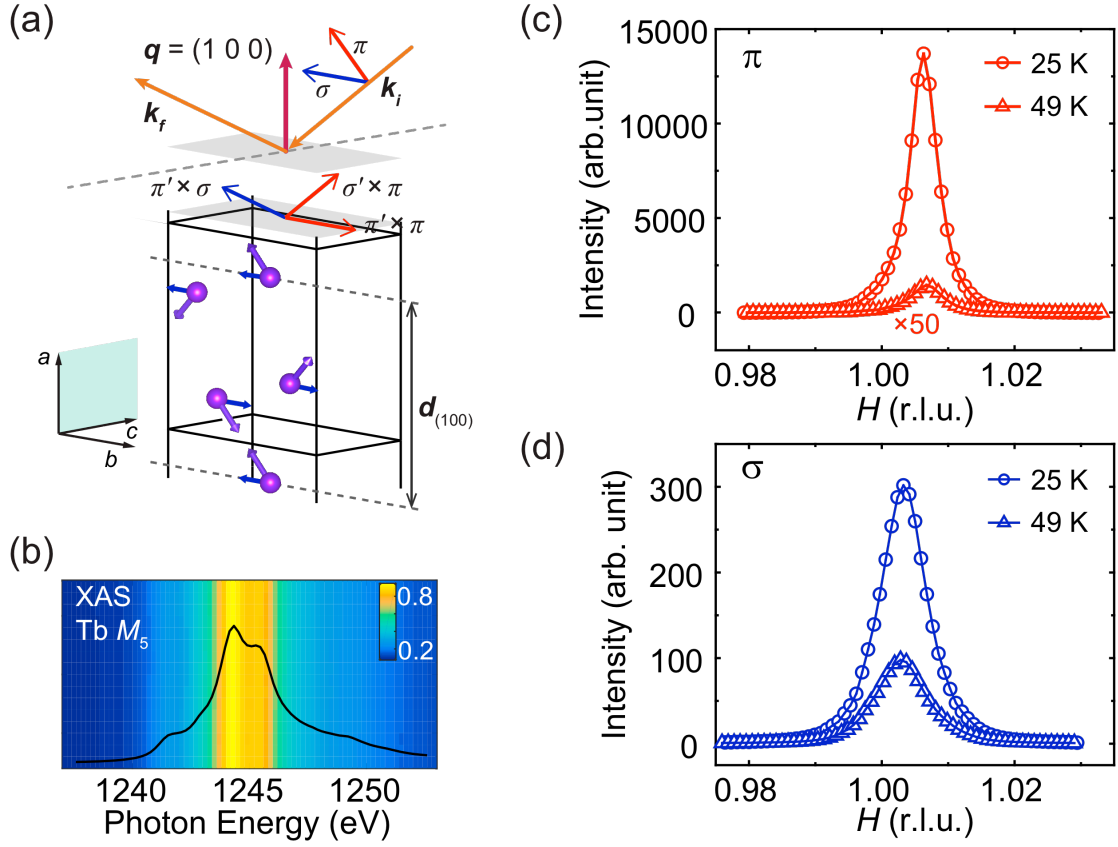

**Figure 2:** (a) The scattering geometry of the RSXS experiment and the polarization configuration. $k_i$ and $k_f$ represent the incident x-ray and scattering x-ray, respectively. The blue arrows represent the spin component along *b*-axis. The $d_{(100)}$ denotes to the real space periodicity of the (1 0 0) magnetic reflection. (b) The XAS spectrum of $TbB_4$ around the Tb $M_5$-edge. (c, d) The (1 0 0) peaks in incident $\pi$ channel and $\sigma$ channel. The triangles and circles indicate temperature above and below $T_{N1}$, respectively

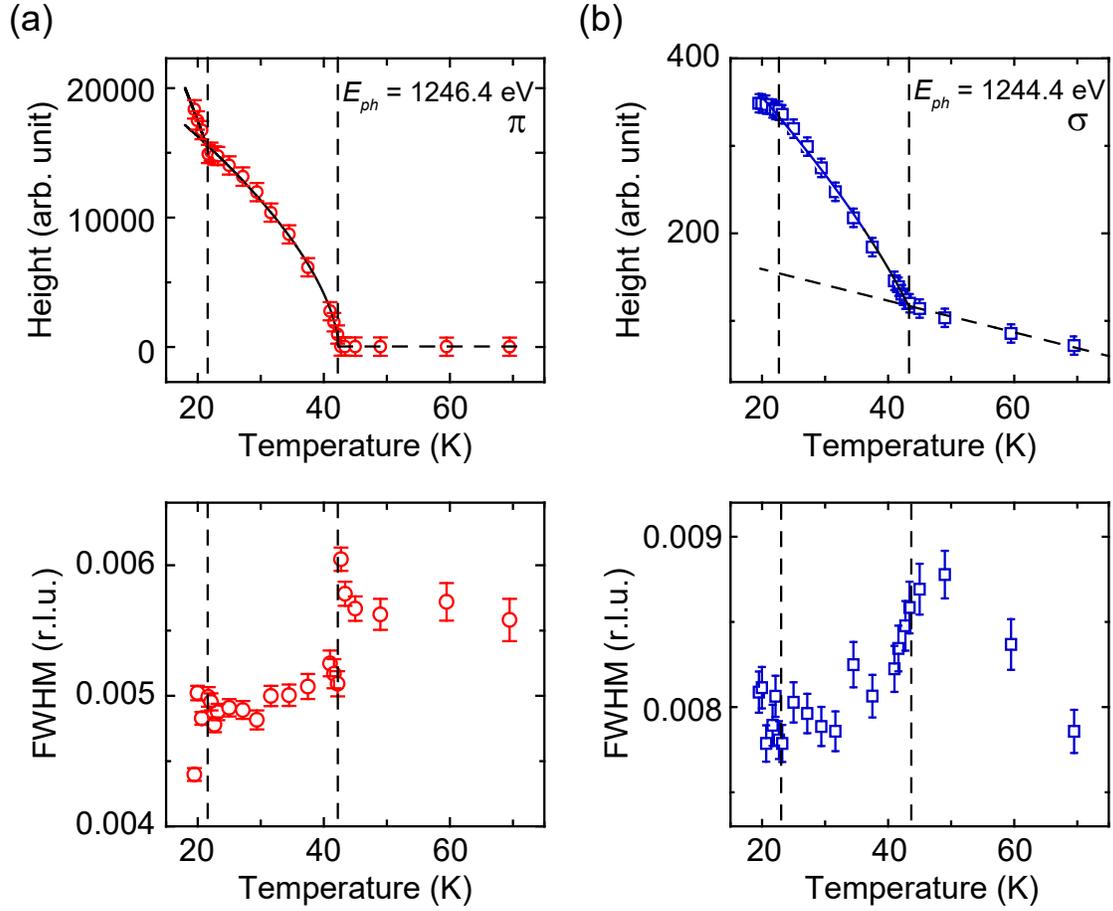

**Figure 3:** The fitting results of the (1 0 0) peaks in both π channel (a) and σ channel (b) as a function of temperature. Top and bottom panels are heights and full-width-half-maximum (FWHM) of the peak fittings by Lorentzian function, respectively. The black curves represent the power law fitting result. Vertical dashed lines represent the phase transition temperature $T_{N1}$ and $T_{N2}$.

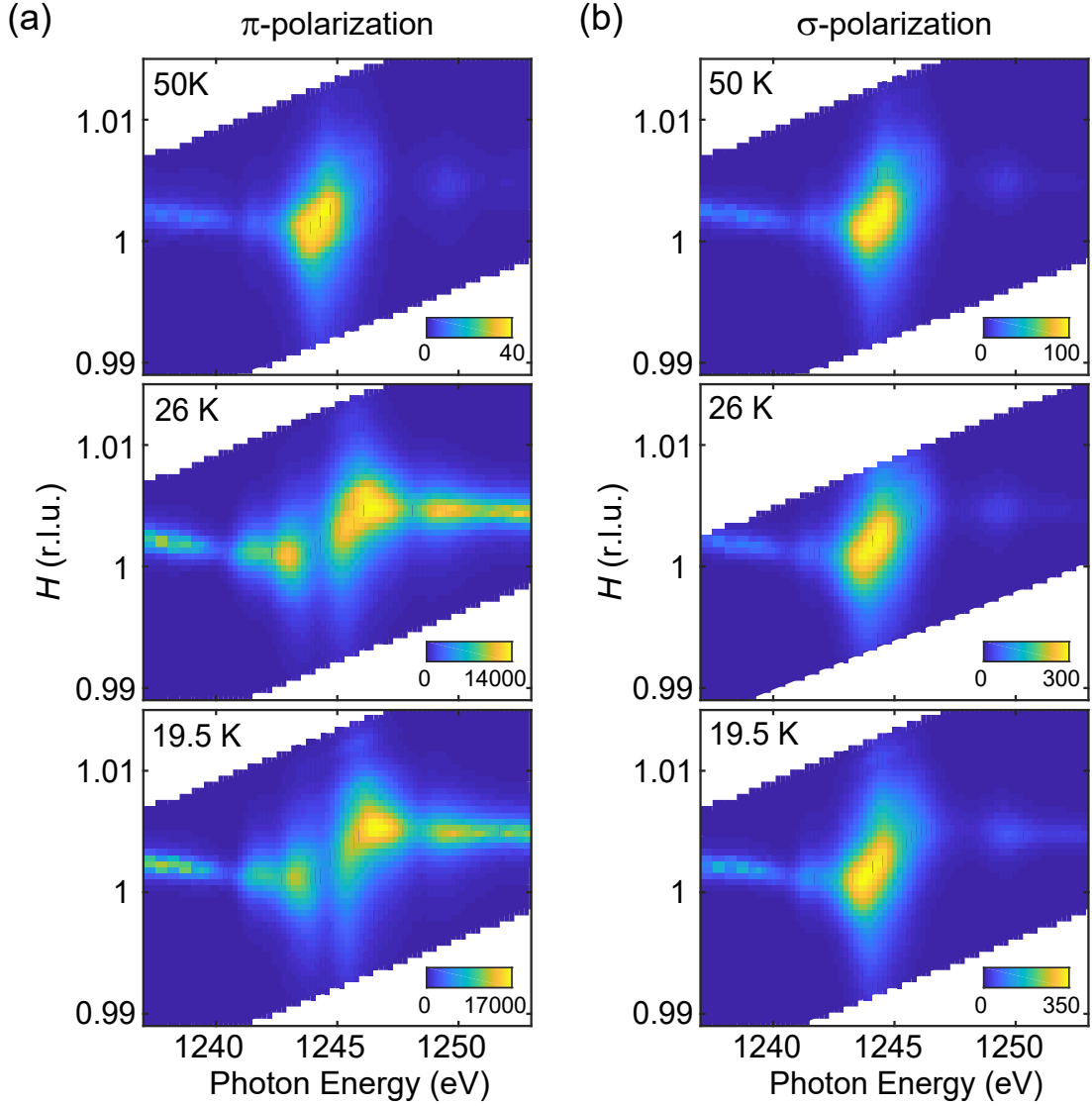

**Figure 4:** Two-dimensional $E_{ph}$-$H$ maps of the (1 0 0) resonant profiles in the π (a) and σ (b) channels. Top, middle, and bottom panels were measured at $T = 50$ K (above $T_{N1}$), 26 K ($T_{N2} < T < T_{N1}$), and 19.5 K (below $T_{N2}$), respectively. The map at both π and σ channels with $T = 50$ K indicates the ATS reflections.

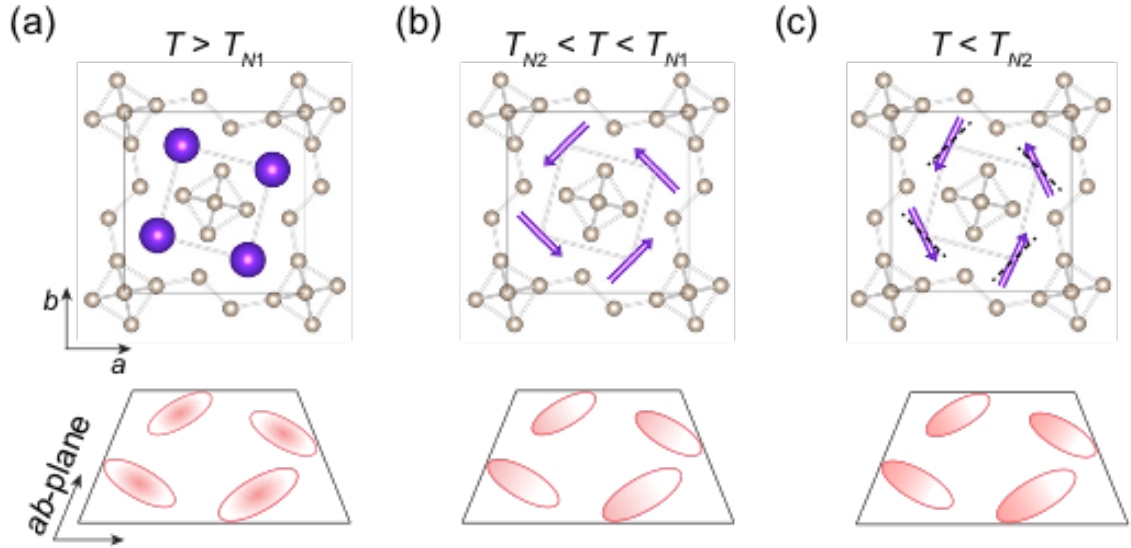

**Figure 5:** Schematic picture of the evolutions of both Tb's spin (arrows) and orbital (ellipses) orders in the crystal *ab*-plane as a function of temperature. (a) At $T > T_{N1}$, no AFM order develops yet. The forbidden ATS reflection exists. (b) At $T_{N2} < T < T_{N1}$, AFM order turns on while QO is developing. (c) At $T < T_{N2}$, the in-plane Tb spins are rotating while QO is staying. The dashed lines denote the spin configuration at $T_{N2} < T < T_{N1}$.

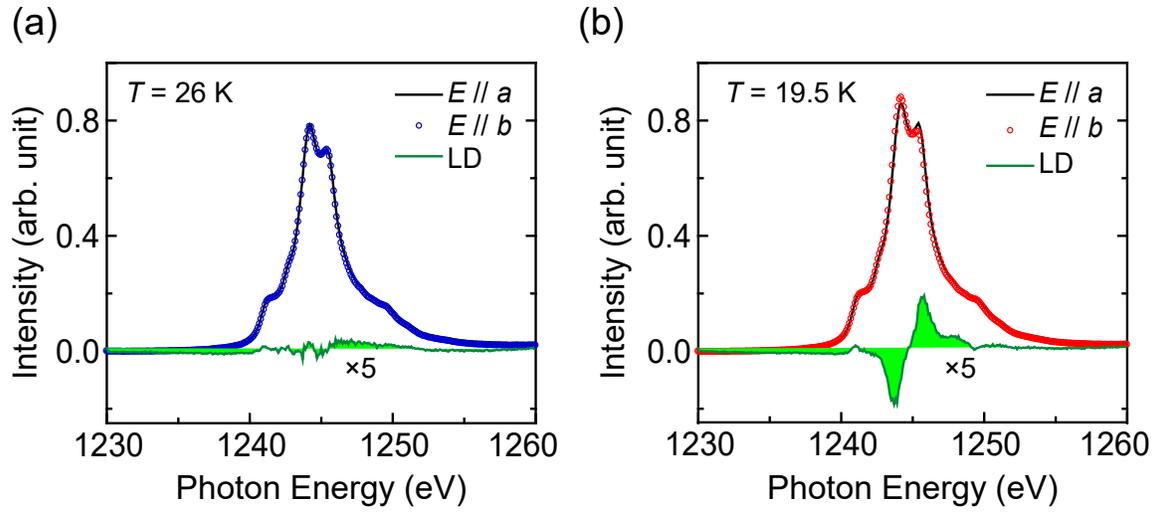

**Figure 6:** Polarization dependence of the XAS on TbB$_4$ at (a) $T = 26$ K (above $T_{N2}$) and (b) $T = 19.5$ K (below $T_{N2}$). The incident photon polarizations are aligned at either the crystal *a*-axis ($E$ // $a$) or the crystal *b*-axis ($E$ // $b$). The LD signal is extracted by $E$//$a$ − $E$//$b$.